\documentclass[12pt,preprint]{aastex}

\begin{document}

\title{The two sided parsec scale structure of the Low Luminosity
  Active Galactic Nucleus in NGC~4278}
\author{M. Giroletti\altaffilmark{1,2},
        G. B. Taylor\altaffilmark{3,4},
	and G. Giovannini\altaffilmark{1,2}}

\altaffiltext{1}{Istituto di Radioastronomia, CNR/INAF, via Gobetti
  101, 40129, Bologna, Italy}
\altaffiltext{2}{Dipartimento di Astronomia, Universit\`a di Bologna,
  via Ranzani 1, 40127 Bologna, Italy}
\altaffiltext{3}{Kavli Institute of Particle Astrophysics and
  Cosmology, Menlo Park, CA 94025, USA}
\altaffiltext{4}{National Radio Astronomy Observatory, P.O. Box O,
  Socorro, NM 87801, USA}


\begin{abstract}

We present new Very Long Baseline Interferometry observations of the
LINER galaxy NGC~4278. The observations were taken with the Very Long
Baseline Array (VLBA) and a single antenna of the Very Large Array
(VLA) at 5 GHz and 8.4 GHz and have a linear resolution of $\la
0.1$~pc. Our radio data reveal a two sided structure, with symmetric
$S$-shaped jets emerging from a flat spectrum core. We fit the jet
brightness with gaussian components, which we identify from a previous
observation taken five years before. By comparing the positions of the
components in the two epochs, we measure motions between $0.45 \pm
0.14$ and $3.76 \pm 0.65$ mas, corresponding to apparent velocities
$\la 0.2\,c$, and to ages in the range $8.3 - 65.8$~years. Assuming
that the radio morphology is intrinsically symmetric and its
appearance is governed by Doppler beaming effects, we find that
NGC~4278 has mildly relativistic jets ($\beta \sim 0.75$), closely
aligned to the line-of-sight ($2^{\circ} \le \theta \le
4^{\circ}$). Alternatively, the source could be oriented at a larger
angle and asymmetries could be related to the jet interaction with the
surrounding medium. We also present new simultaneous VLA observations
between 1.4 and 43 GHz, and a 5 GHz light curve between 1972 and
2003. The radio spectrum can be fit by a relatively steep power-law
($\alpha = 0.54$).  We find significant variability at 5~GHz. All
these arguments indicate that the radiation from NGC~4278 is emitted
via the synchrotron process by relativistic particles accelerated by a
supermassive black hole. Despite a much lower power, this is the same
process that takes place in ordinary radio loud AGNs.

\end{abstract}

\keywords {galaxies: active --- galaxies: nuclei --- galaxies:
individual (NGC~4278)}

\section{Introduction}

Although objects hosting an Active Galactic Nucleus (AGN) represent
only a small fraction of the total number of extragalactic sources,
there is growing evidence that a low level of nuclear activity is a
common feature among galaxies. Objects presenting a spectral signature
of such activity include low-ionization nuclear emission-line region
\citep[LINERs, ][]{hec80}, low luminosity Seyfert galaxies, and
``transition nuclei'', i.e. nuclei with spectra intermediate between
LINERs and HII regions. These objects are grouped under the name of
low luminosity active galactic nuclei \citep[LLAGN, see][]{ho97}.

Dynamical searches and stellar kinematics have yielded strong evidence
for the existence of compact dark objects of mass $\sim 10^6$ to
$10^{9.5} M_\sun$ \citep{kor95,mag98} in virtually all bulge-dominated
galaxies. These are most likely in the form of super massive black
holes (SMBHs); however, what is still unknown is the state of the
SMBH, and the reason why they are so often quiescent. In the light of
these findings, the debate over the physical processes that are
responsible for the ``activity'' in LLAGN has been a major issue in
recent years. These galaxies may host a scaled down version of the
more powerful AGN, with an accreting super massive black hole rendered
under-luminous because of a low accretion rate and/or low radiative
efficiency.  Alternatively, the activity might be related to stellar
processes \citep[e.g.,][]{shi92,fil92,alo00}.

Along with hard X-ray data \citep[e.g.][]{ho01,ter02}, radio
observations are a useful tool to discriminate between an AGN and a
starburst powered nucleus. VLA observations with arcsecond
\citep{fil00,fil02} or subarcsecond \citep{nag00,nag02} resolution
have successfully revealed the presence of a compact radio core in
most LINERs and low luminosity Seyferts, with the non-detections
likely caused by limits in sensitivity only. The corresponding
brightness temperatures are $T_b \ge 10^{2.5 - 4}$ K, which are
consistent with a non-thermal origin. The spectral index of these
radio cores is typically flat -- a well-known signature of synchrotron
self-absorbed emission \citep{nag01}.

High angular resolution Very Long Baseline Interferometry (VLBI)
observations yield more conclusive evidence: all (16/16) LLAGNs in a
flux density and distance limited sample possess milliarcsecond
compact cores, as revealed with the VLBA at 6 cm
\citep{fal00,nag02}. This implies brightness temperatures $T_B \ga
10^8$ K, which definitely rule out any explanation based on thermal
processes, and point directly to synchrotron emission from the base of
a parsec scale jet. Indeed, the five sources with the brightest cores
have parsec scale jets.  Furthermore, high frequency radio spectra can
be used to discriminate between pure ADAF \citep[Advection Dominated
Accretion Flows,][]{nar95} models and non-thermal jets
\citep{dim99,dim01,and04}.


The nearby elliptical galaxy NGC~4278 has been investigated in detail
at most wavebands. In the optical, HST observations reveal a central
point source and a large distribution of dust located north-northwest
of the core \citep{car97}. Ionized nuclear gas typical of a LINER is
found in this galaxy \citep{gou94}, possibly associated with an
external ring of neutral hydrogen in PA 135$^{\circ}$ \citep{rai81}.

Radio continuum observations of NGC~4278 on kpc scale have been
carried out at frequencies between 5 GHz and 43 GHz, revealing a
compact source \citep{dim01,nag01,nag00}. At 1.4 GHz, both the FIRST
\citep{bec95} and the NVSS \citep{con98} data confirm that the source
is compact. Only at 8.4 GHz, at a resolution of $\sim 200$ mas, is
NGC~4278 slightly resolved into a two-sided source with an extension
to the south \citep{wil98}. Compactness and a flat radio spectrum
suggest that the radio emission is non-thermal, and in this respect
the emission from NGC~4278 seems to be very similar to that of
powerful radio loud AGNs, such as QSO and BL Lacs; however, the total
radio luminosity of the source is only $P_\mathrm{1.4\, GHz} =
10^{21.6}$ W Hz$^{-1}$, i.e. at least two orders of magnitude less
than those powerful objects.

On parsec scales, early VLBI experiments at 18 cm and 6 cm have
revealed a core dominated structure, with an elongated feature
extending to the north-west and possibly to the south on scales of
some 10 mas \citep{jon81,jon82,sch83,jon84}. More recent observations
with the VLBA at 6 cm have yielded a dramatic improvement in our
knowledge of this source: \citet{fal00} reveal an extended core and an
elongated region to the southeast on scales of a few milliarcsecond;
\citet{gio01}, thanks to more short spacings provided by a VLA antenna
in addition to the full VLBA, have detect emission on the opposite
side of the core as well. However, the discrepancy in the total flux
density [$(87 \pm 4)$ vs. $(400 \pm 20)$~mJy] is more than the simple
addition of a VLA antenna can justify; it is possible that the source
may be variable or simply that there have been problems in the
calibration of the flux density scale. \citet{bon04} also detect a trace of
two-sided emission, although heavily resolved.  Finally, VLBA
phase-referenced observations have succeeded in detecting the source
on sub-pc scale even at 43 GHz, where it shows only a core and a hint
of low level emission to the north \citep{ly04}.

In the present paper, we consider new VLBA observations at two
frequencies (5 GHz and 8.4 GHz), taken on 2000 August 27. We compare
the new 5 GHz image to previous images and discuss the morphology and
the motion of components. Furthermore, the 8.4 GHz data provide a
favorable combination of resolution and sensitivity, constraining the
brightness temperature and the location of the central black hole.

We describe the new observations in \S 2 and present the results in \S
3. A discussion is given in \S4, and the main conclusions are
summarized in \S5. Throughout this paper, we adopt for the Hubble
constant a value of 71 km s$^{-1}$ Mpc$^{-1}$; however, since NGC~4278
has a direct distance measurement of 14.9 Mpc \citep{ton01,jen03}, we
do not make use of its redshift ($z=0.00216$), which is affected by
proper motion in a galaxy group. At the distance of NGC~4278, 1 mas
corresponds to a linear scale of 0.071 pc.  We define the spectral
index $\alpha$ following the convention that $S(\nu) \propto
\nu^{-\alpha}$.

\section{Observations and Data Reduction}

\subsection{VLBA observations}
We observed NGC~4278 with an 11 element VLBI array composed of the
NRAO Very Long Baseline Array (VLBA) and a single 25~m VLA antenna for
10 hours. The observing run was performed on 2000 August 27, switching
between 5 GHz (with 4 IFs at 4971.49, 4979.49, 4987.49, 4995.49 MHz)
and 8.4 GHz (8405.49, 8413.49, 8421.49, 8429.49 MHz). Sixteen channels
per IF and full polarization data were recorded. Short scans on 3C279
were taken in order to find fringes, and several observations of OQ208
were repeated at different hour angles in order to calibrate the
polarization leakage terms. J1310+3220 and J1751+0939 were also
observed to perform checks on the amplitude and polarization
calibration.

The correlation was carried out at the AOC in Socorro. The
distribution tapes were read into the NRAO Astronomical Image
Processing System (AIPS) for the initial calibration.  We followed the
same scheme for the data reduction of both 5 GHz and 8.4 GHz
data-sets. As a first step, we corrected our data using the accurate
position information obtained by \citet{ly04} in a phase-referenced
experiment with the VLBA at 43 GHz (RA 12$^h$ 20$^m$ 06$^s$.825429,
Dec $29^{\circ}$ $16\arcmin$ $50\arcsec.71418$). We then performed the
usual calibration stages (removal of instrumental single band delay,
and the $R-L$ delay, and bandpass calibration) using scans on 3C279.

After flagging bad data, we obtained good models for the calibrators,
which we used to improve the amplitude calibration for the entire
data-set. Finally, we removed the instrumental polarization leakage
using OQ208. A single-source file for NGC~4278 was then created by
averaging the frequency channels and IFs and exported into Difmap
\citep{she94,she95} for editing and self-calibration.

\begin{figure}
\plotone{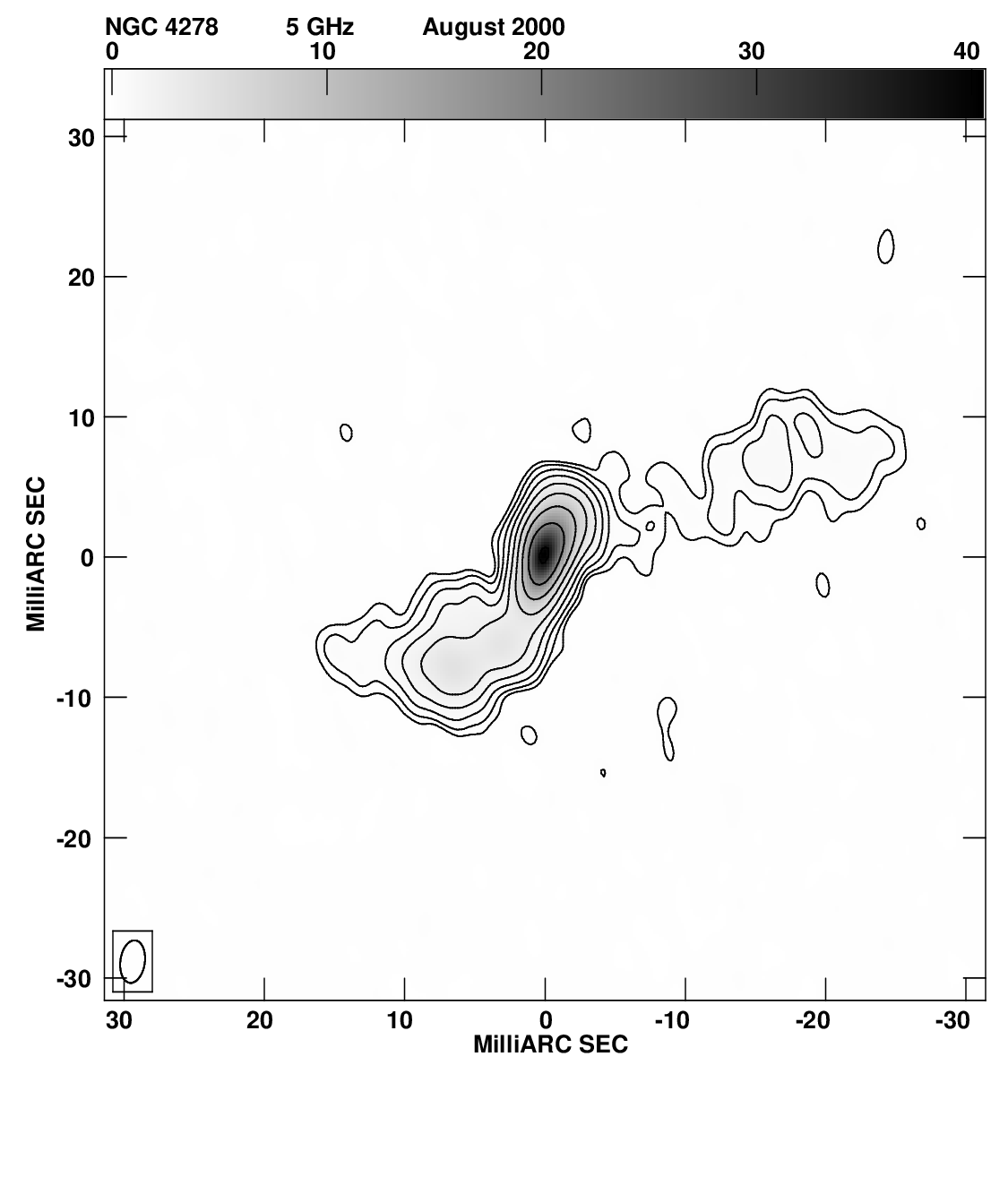}
\caption{VLBA+Y1 image of NGC~4278 at 5 GHz, 2000 August 27. Contours are
  drawn at (1, 2, 4, ..., 128) times the lowest contour, which is
  0.150 mJy/beam. The grey scale range is from $-0.2$ to 40.5 mJy/beam.
  \label{fig1} }
\end{figure}

\begin{figure}
\plotone{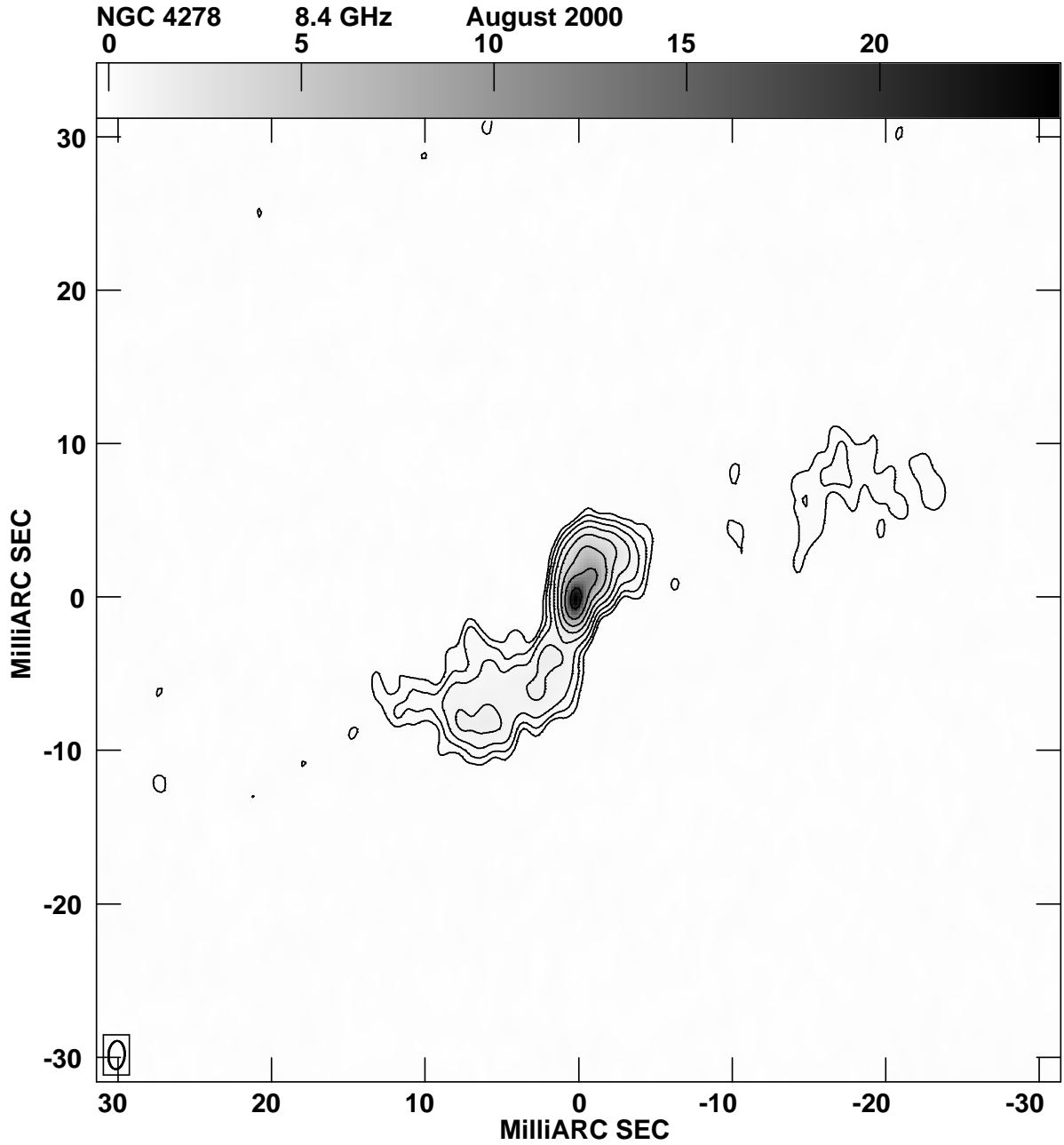}
\caption{VLBA+Y1 image of NGC~4278 at 8.4 GHz, 2000 August 27. Contours are
  drawn at (1, 2, 4, ..., 128) times the lowest contour, which is
  0.150 mJy/beam. The grey scale range is from $-0.2$ to 24.6 mJy/beam.
  \label{fig2} }
\end{figure}

Thanks to a good calibration and position information, we could obtain
final images with only a few iterations of phase self-calibration.
One cycle of amplitude self-calibration with a long solution interval
(30 minutes) has also been performed before obtaining the final
$(u,v)-$data. Total intensity images with natural weights are
presented in Figs.~\ref{fig1} and \ref{fig2}, and the most relevant
parameters are summarized in Table~\ref{tab1}. Stokes 'Q' and 'U'
images were also made.

The final calibrated datasets are not suitable for a direct
combination into a spectral index map. Different observing frequencies
correspond to different coverage of the $(u,v)$-plane and this could
introduce spurious results in a spectral index map. Therefore we
selected an annulus in the $(u,v)$-plane, consisting of only baselines
in the range 0.9 M$\lambda - 145$ M$\lambda$. We also tapered the data
at 100 M$\lambda$ in $u$ and 60 M$\lambda$ in $v$ in order to account
for the more elliptical coverage of the 5 GHz data.  After this
processing, the images have a resolution of 3.2 mas $\times$ 1.9 mas
in PA $-2^{\circ}$ and were combined to produce the spectral index map
using the AIPS task {\tt COMB}.

We also re-analyzed VLBA+Y1 5 GHz data obtained in 1995 (July 22),
taking advantage of the new position \citep{ly04}. The position used
in the original correlation was off by more than one arcsecond from
the actual position of the source. This resulted in exceedingly large
delays which caused a lot of failures in the fringe fitting
process. The new accurate positional information allows us to recover
a large fraction of useful data, resulting in a significantly
improved, high fidelity image. We present and briefly discuss the
results from the reprocessed data in Sect.~\ref{jul95}.

\subsection{VLA observations}
New VLA observations were taken at 1.4, 5.0, 8.4, 15, 22, and 43 GHz
in A configuration on 2003 August 17 and reduced in AIPS following
standard procedures. 3C\,286 was used to set the flux density
scale. Phase referencing to the nearby calibrator 1221+282 failed at
high frequencies due to bad weather (rain).  Thus, we only have an
upper limit at 43 GHz and the value at 22 GHz is more uncertain than
at lower frequencies -- we estimate a fractional error of 2\% up to
8.4 GHz, of 5\% at 15 GHz, and of 10\% at 22 GHz.

\section{Results}

\subsection{VLBA on 27 August 2000}

The final images reveal a source dominated by a central compact
component, with emission coming from either side. To the southeast, a
jet-like feature extends for $\sim 6.5$ mas in PA 155$^{\circ}$
(measured north to east), then progressively bends into PA
100$^{\circ}$. In total, the jet is almost 20 mas long, which
corresponds to $\sim 1.4$ pc. On the opposite side, the main component
is slightly elongated to the north in the 5 GHz map, and the 8.4 GHz
data clearly show a secondary component in PA $-40^{\circ}$. Then,
this jet-like feature bends to the west turning into a diffuse,
uncollimated, low brightness region.

In total, the source extends over $\sim 45$ milliarcsecond, i.e. about
3 parsecs. Given its morphology and dimension, NGC~4278 meets all
requirements for the classification as a Compact Symmetric Object
(CSO), except that it is under-luminous compared to most known CSOs
\citep{rea94,pec00}.

The total flux density measured in our images is 120 mJy at 5 GHz and
95 mJy at 8.4 GHz. If we compare these values with those obtained with
the VLA (162 mJy and 114 mJy, respectively, see Table~\ref{tabvla}),
we find about a 20-25\% difference, which can be ascribed to the VLBA
resolving out some extended emission, probably in the western
region. The monochromatic luminosity at 1.4 GHz is $3.18 \times
10^{21}$ W Hz$^{-1}$, and $2.52 \times 10^{21}$ W Hz$^{-1}$ at 8.4
GHz.

\begin{figure}
\plotone{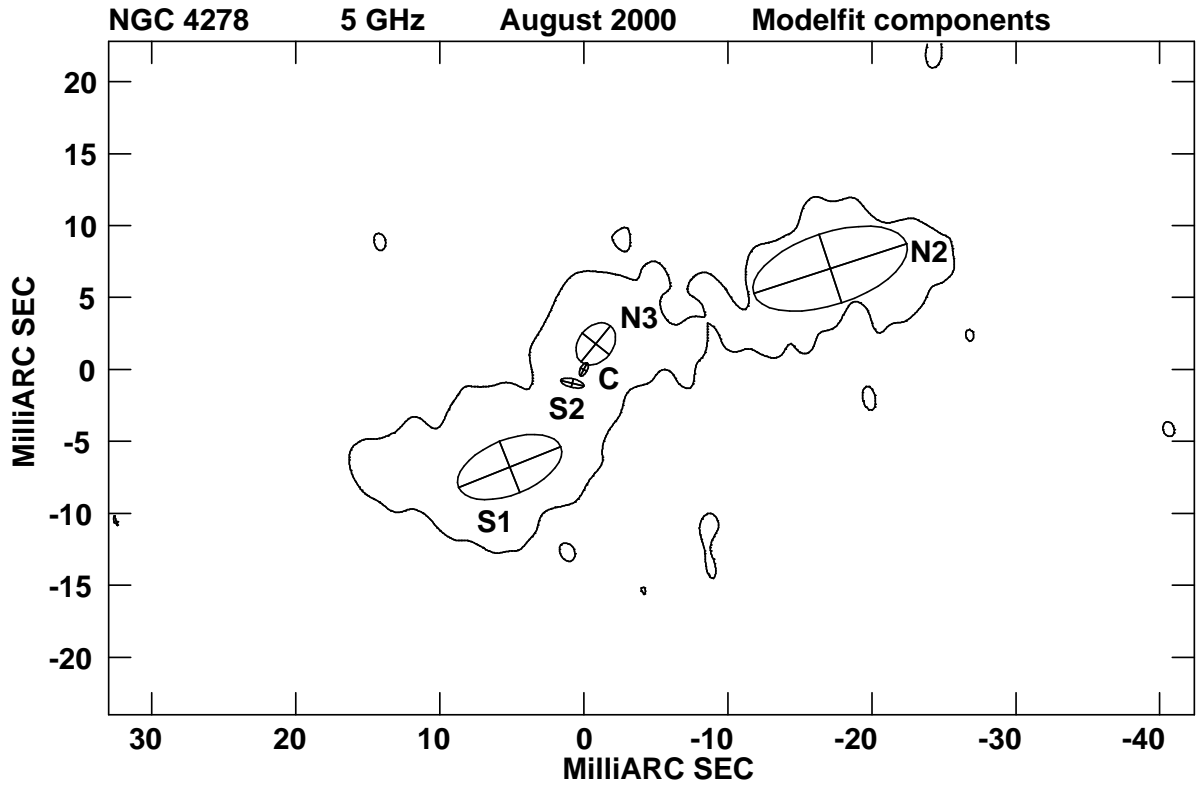}
\caption{Model components for epoch 2000.652, overlaid on the lowest
  contour from the 5 GHz image (150 $\mu$Jy/beam). Parameters of the
  components are given in Table~\ref{tab2}. \label{figcomp} }
\end{figure}

The visibility data are well fit by a five component model at both
frequencies. The position and dimension of the components are
illustrated in Fig.~\ref{figcomp}; our choice for labeling the
components is based on their most likely epoch of ejection, as
discussed in \S~\ref{motion}. Table~\ref{tab2} shows the relevant
parameters for all components, re-referenced to the central component
$C$. To compute positional uncertainties for the components, we
repeated the modelfit on the 'LL' and 'RR' correlations independently;
the semi-dispersion of the results is taken as a $1 \sigma$ error on
the position of each component. This yields $\Delta r = 0.04$~mas for
components $S2$ and $S1$, 0.01~mas for $N3$ and 0.34~mas for $N2$.  In
Columns 7 and 8 we report the flux density at 5 and 8.4 GHz
respectively, and in Column 9 we compute the spectral index. Besides
being the most compact feature, component $C$ presents also the
flattest spectral index ($\alpha = 0.2$) and its identification with
the core is straightforward.

Finally, no polarized signal is detected from our data. Both the
images in Stokes 'Q' and 'U' are purely noise-like, with on source $1
\sigma$ noise levels of 66 and 63 $\mu$Jy/beam at 5 GHz and of 62 and
60 mJy/beam at 8.4 GHz, respectively. Based on the on-source noise in
the 'Q' and 'U' images, we place a limit on any polarized signal for
the source at $< 125 \ \mu$Jy/beam at 5 GHz and $<120 \ \mu$Jy/beam at
8.4 GHz ($3\sigma$ values). At the position of the brightest component
in the total intensity images, this corresponds to a fractional
polarization of $< 0.3\%$ at 5 GHz and $< 0.5\%$ at 8.4 GHz.

\subsection{VLBA on 22 July 1995}
\label{jul95}
\begin{figure}
\plotone{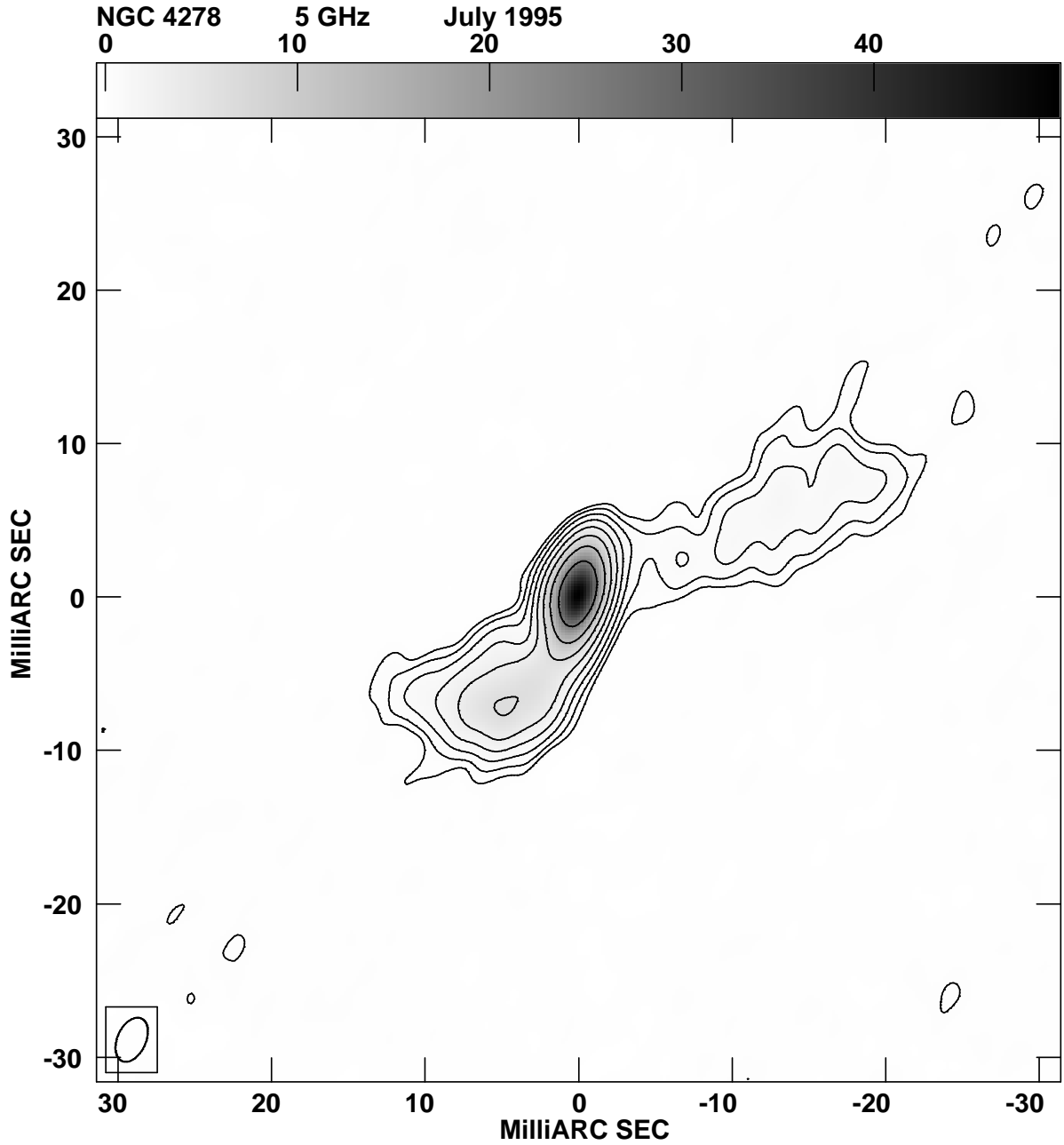}
\caption{VLBA+Y1 image of NGC~4278 at 5 GHz, 1995 July 22. Contours are
  drawn at (1, 2, 4, ..., 128) times the lowest contour, which is
  0.195 mJy/beam. The grey scale range is from $-0.3$ to 49.5 mJy/beam.
  \label{fig4} }
\end{figure}

The reprocessing of the July 1995 data has yielded a significant
improvement over the previously published results. The new image
(Fig.~\ref{fig4}) is in good agreement with the data taken at 5 GHz in
2000, showing a central compact component and jet-like features in the
same position angle. The overall morphology of the source is similar
to that shown in \citet{gio01} and we confirm the reality of the
emission coming from the northwest. However, the noise level has
improved by over a factor of 6 (65 vs 400 $\mu$Jy) and the flux
density scale is more consistent with both our new observations as
well as the results presented in \citet{fal00} and \citet{bon04}. The
resulting total flux density is 135 mJy.

The same five component model that fits the 2000 epoch data has been
applied to this data set as well, allowing for the components to
change in flux density and position. The resulting fit is good
(reduced $\chi$-squared of 0.98) and is presented in
Table~\ref{tab3}. Uncertainties are computed in the same way as for
epoch 2000.652, and are 0.13, 0.11, 0.09, and 0.56~mas for $S2$ $S1$,
$N3$, and $N2$, respectively.

An independent fit with only four components has also been tried, and
fits the data almost as well as the five component model (reduced
$\chi$-squared of 1.00): components S2, S1, and N2 are maintained in
both models, while $C$ and $N3$ can actually be fitted with one
Gaussian only ($C'$).  However, this model gives a less accurate
representation of the core region, with the component $C'$ too
extended (3.2 mas $\times$ 1.3 mas) to match the size of the compact
core. We also note that if $N3$ is removed from the model for the
2000.652 epoch data, any following iteration of the fitting algorithm
will reintroduce it at the expense of $S2$.

\subsection{Component motion and variability}
\label{motion}

\begin{figure}
\plotone{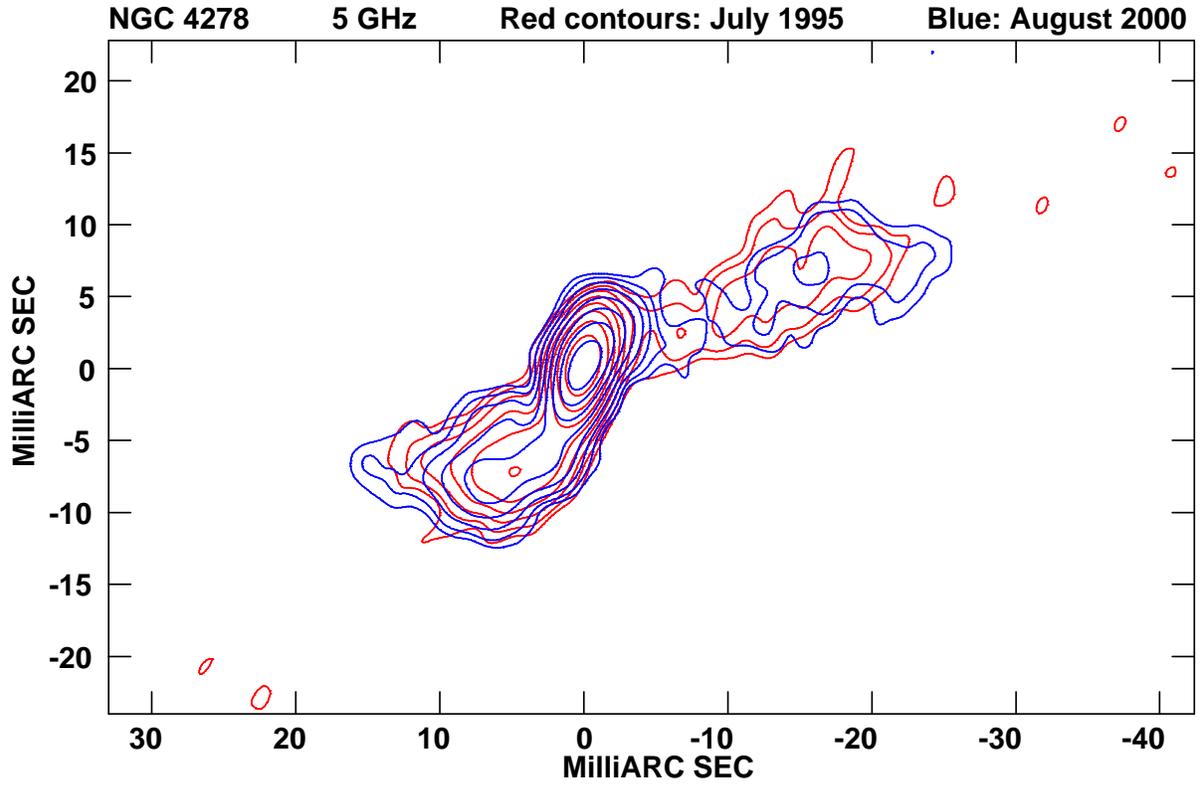}
\caption{An overlay of the intensity contours at 5 GHz for epoch 1995.551
  (red) and 2000.652 (blue). Both images have been convolved with the
  1995.551 beam (3.0 mas $\times$ 1.8 mas in PA $-25^{\circ}$) and
  plotted with the same contours, i.e. (1, 2, 4, ...) $\times$ 0.2
  mJy/beam \label{fig5} }
\end{figure}

If we compare data taken at the same frequency in different epochs, we
can get information on the evolution of the source. A first comparison
can be made by directly overlaying the contours at the same frequency
from the two epochs (Fig.~\ref{fig5}). Before overlaying them, we have
convolved the two images with the same restoring beam. We also plot
the same contours, regardless of the lower noise level of the 2000.652
image. This guarantees that any difference is due to a real change in
the source structure and not simply to the better sensitivity of one
observation.

The overlay is suggestive of a displacement of the northwest low level
emission, moving away from the core. Minor differences can be seen
also in the core region and in the southeastern jet. In general, the
blue 2000.652 contours trace the emission in a region more extended
than the red 1995.551 ones.  This behavior can be better quantified if
we consider the relative motion of the components identified by the
modelfit process. Taking the position of the core (component $C$) as a
reference, and assuming it is fixed, we have compared the position of
the other components. We report the results in Table~\ref{tab4}:
column (1) labels the components, columns (2) and (3) give the
relative motion over the five years in polar coordinates (distance in
mas, angle in degrees); columns (4) and (5) report the apparent
velocity in mas~yr$^{-1}$ and units of $c$, respectively; finally, the
corresponding kinematic age is given in column (6). The radial
distance of each component has increased over the five years lag
between the observations. The motion is larger on the northwestern
side; in particular, the largest displacement is found for component
$N2$. The corresponding apparent velocities are given in Column~(5),
with the fastest component moving at $\sim 0.17\ c$.

Assuming that the apparent velocity is constant for each component, we
derive ages as reported in column~(6). $S2$ and $N2$ have ages that
are consistent, and they could have been ejected together about 25
years before epoch 2000.652. $S1$ is the oldest component, and its
counterpart in the main jet is not detected, probably because it is
too distant and extended. Finally, $N3$ is the youngest component,
ejected only three years before our first epoch of observations; it is
likely that a corresponding component $S3$ has emerged in the
counterjet but is still confused with the core. Note that the core is
the only component whose flux density is larger in 2000.652 than in
1995.551.

Finally, we can compare the total flux density detected in the two
epochs. Our data indicate a variation of the total flux density of
about 10\%, with the source being less powerful in the second
epoch. Note that \citet{fal00} report a value of 87.3 mJy; however,
that observation was performed {\it without} the addition of a VLA
antenna, which results in a lack of short spacings sensitive to low
level emission. As a consequence, component $N2$ and (partly) $S1$ are
resolved out, accounting for the difference in total flux density.

\subsection{Spectral index}
\label{spix}

\begin{figure}
\plotone{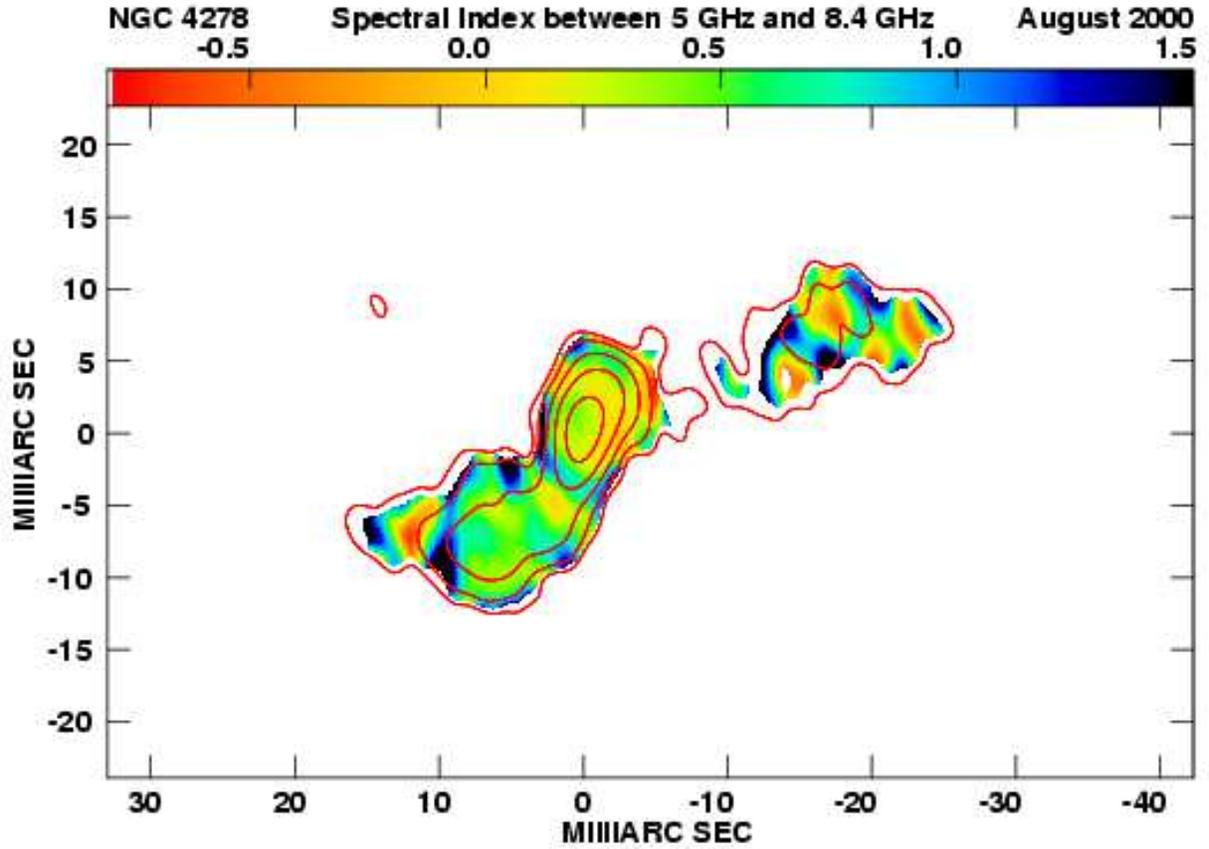}
\caption{The spectral index map between 5 GHz and 8.4 GHz with overlaid
  contours at 5 GHz. The color scale range is between $-$0.8 (red,
  inverted indices) and 1.5 (blue, steep spectrum); contours are drawn
  at (1, 3, 10, 30, 100) times the lowest contour, which is 0.2
  mJy/beam. In our convention, $S(\nu) \propto \nu^{-\alpha}$.
  \label{fig6} }
\end{figure}

We now compare data taken at the same epoch but with different
frequencies, in particular we present the map of the spectral index
from 2000 August 27 (Fig.~\ref{fig6}). The features of the spectral
index map are in agreement with the single component values determined
by the model-fit. The core region presents the flattest spectrum
($\alpha \sim 0.0$), steepening along the southern jet ($\alpha \ga
0.5$). An irregular spectral index pattern is found in regions with
lower level emission, such as the outermost edge of the southern jet
and the extended northern component N2. However, we can still
determine an average spectrum for this region, which has $\alpha =
0.8$.

\begin{figure}
\plotone{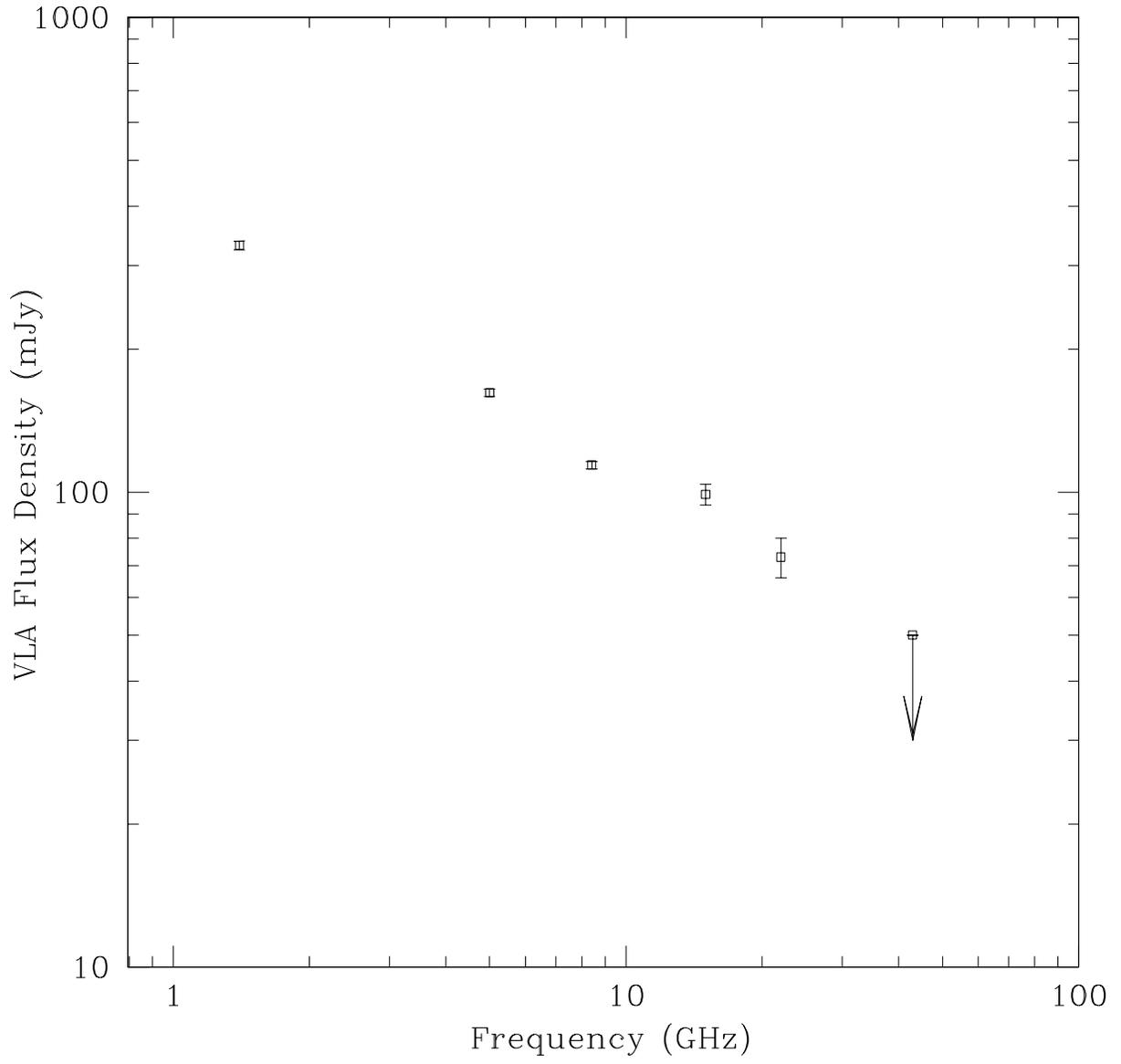}
\caption{VLA flux density measurements between 1.4 GHz and 43
  GHz. \label{figvla} }
\end{figure}

In Fig.~\ref{figvla} we also plot the simultaneous spectrum taken with
the VLA in A configuration from 1.4 GHz to 43 GHz. The measured flux
densities are reported in Table~\ref{tabvla}. The best-fit power-law
spectral index is $\alpha = 0.54$.

\section{Discussion}

\subsection{Jet orientation and velocity}
\label{jets}

Our images detect low level emission to the northwest of NGC~4278 with
unprecedented resolution and sensitivity.  We find, in the inner part
of the jet, a compact component N3, and at a larger distance, $\sim
10$ mJy of flux density in the region N2. Thus, we classify NGC~4278
as a two-sided source, similarly to a few other LLAGN previously
studied, e.g. NGC~4552 \citep{nag02}, NGC~6500 \citep{fal00}, and
NGC~3894 \citep{tay98}.

We are therefore faced with the puzzle of determining which of the
jets is pointing toward us and the direction of the source main axis
with respect to the line-of-sight. Although the southern jet looks
more collimated, the total flux density is larger in the northern
components than in the southern ones. The high resolution 8.4 GHz
image (Fig.~\ref{fig2}) clearly shows that the inner jet is brighter
to the north than to the south. The apparent motion of the northern
components are also larger than those of the southern ones. Finally,
the 43 GHz image in \citet{ly04} shows a hint of emission only to the
northwest of the core on scales $<$ 1 mas. We therefore argue that the
main, approaching jet is the northern one.

\subsubsection{Jet Asymmetries from Doppler Boosting}
\label{doppler}

To estimate the orientation $\theta$ and intrinsic velocity $\beta$ of
the jet, we initially consider a simple beaming model, which assumes
that pairs of components are ejected simultaneously from the core with
the same intrinsic velocity and brightness.  We apply this model to
the component pair $N2$/$S2$, which has been ejected simultaneously
according to our motion measurements (\S~\ref{motion}), and we further
assume that $N2$ is the one on the approaching side. In this model,
the ratio between the arm length $r$ and the proper motion $\mu$ of
the two components are related by


$$R = \frac{\mu_{N2}}{\mu_{S2}} = \frac{r_{N2}}{r_{S2}} = \frac{1+\beta\cos\theta}{1-\beta\cos\theta}$$

\begin{figure}
\plotone{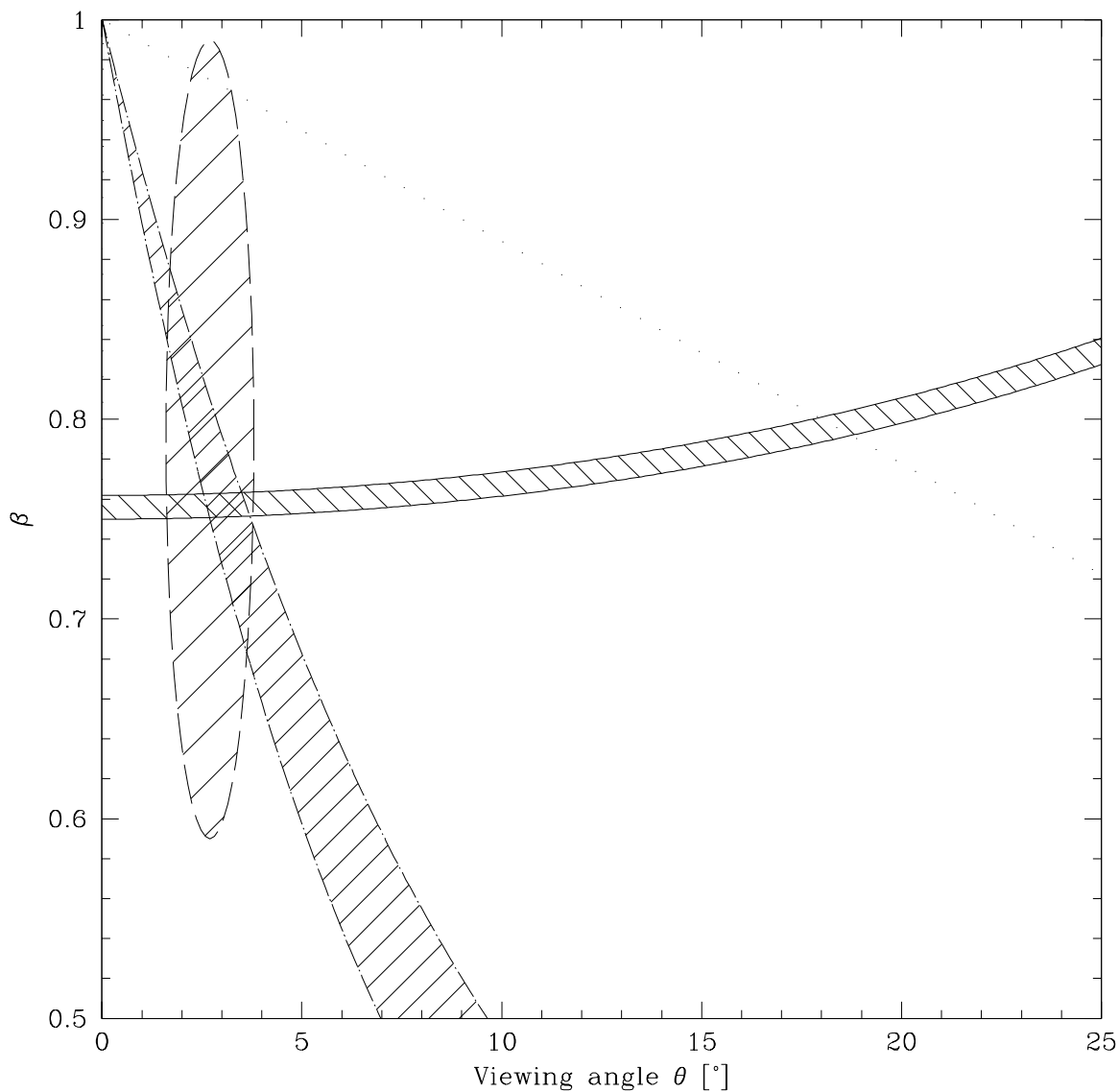}
\caption{$(\theta, \beta)$-plane for NGC~4278, as derived from the
  component pair $N2/S2$. The solid lines are the limits from the arm
  length ratio ($r_{N2}/r_{S2} = 7.2 \pm 0.2$, see discussion), while
  the dot-dash lines represent the results from the apparent
  separation velocity ($\beta_\mathrm{sep}= 0.191 \pm 0.030$);
  finally, the motion ratio and the measured linear distance of the
  source select the region delimited by the dashed ellipse. \label{bt}}
\end{figure}

From our modelfits we derive that $r_{N2}/r_{S2} = 7.2 \pm 0.2$ and
$\mu_{N2}/\mu_{S2} = 8 \pm 3$, so we estimate that $4 < R < 10$. The
arm length ratio corresponds to selecting the hatched area between the
two solid lines in the $(\beta, \theta)$-plane shown in
Fig.~\ref{bt}. The dot-dash lines represent the possible combination
of $\beta$ and $\theta$ resulting from the apparent separation
velocity of the two components, which is expressed by the relation
$\beta_\mathrm{sep}= (2 \beta \sin \theta)/(1-\beta^2 \cos^2
\theta)$. Finally, since we measure motion on both sides and we know
the source distance, we can directly solve for $\theta$ and $\beta$
\citep[see e.g.][]{mir94}; this corresponds to the dashed ellipse
centered on $\theta=2.7^{\circ}$, $\beta=0.79$.

In principle, one could also consider the brightness ratio between the
two components; however, as an effect of relativistic time dilation,
we are watching the components at different stages of evolution. Since
we know little about the time evolution of the jet components, this
hinders the application of the brightness ratio constraint; in any
case, a check that $S_{N2}/S_{S2} > 1$ is possible and is consistent
with our interpretation.

Based on the above analysis (Fig.~\ref{bt}), we find mildly
relativistic velocities of $\beta \sim 0.76$ ($\Gamma \sim 1.5$), and
an orientation close to the line-of-sight ($2^{\circ} \la \theta \la
4^{\circ}$). The resultant Doppler factor is $\delta \sim 2.7$ and
the small viewing angle explains the bends visible in both jets as the
amplification caused by projection effects of intrinsically small
deviations, which are common in high power radio sources.

\subsubsection{Jet Asymmetries from Interactions}

As an alternate explanation to that proposed above, it is possible
that the true radio morphology of NGC 4278 is influenced by more than
relativistic beaming. For example, there may be significant
interactions between the jet and the surrounding medium. This could
explain the lower velocity for the components in the southern side of
the jet and cause the flux density of $S1$ to be enhanced. If this is
the case, the task of identifying corresponding pairs of components
either side of the core becomes confused. The radio source could then
be oriented at a larger angle to the line-of-sight than we deduce from
beaming arguments.  In this scenario, the large bends visible in our
images would be real and not amplified by geometrical effects,
suggesting a strong interaction of the jets with the surrounding
medium. The combination of a strong interaction with a dense medium
and a low power core could well account for the small size of
NGC~4278. It is unlikely, in any case, that more than mildly
relativistic velocities are present in the jet of NGC~4278.

It is interesting to note that \citet{nag02} have detected compact
cores in LLAGNs almost exclusively in Type 1 objects. Among the
possible explanations, they invoke the unified scheme, assuming that
all LLAGNs have accreting black holes and that we only see radio
emission in Type 1 objects because of beaming and/or obscuration. Our
resulting viewing angle determined in \S\ref{doppler} seems to support
this picture; however, the Lorentz factor is not as high as in more
powerful classic radio galaxies, where $3 \la \Gamma \la 10$
\citep{gio01}. Therefore, a source similar to NGC~4278 but oriented at
$60^\circ$, would have a similar Doppler factor $\delta \sim 1.1$,
hence still be detectable. We note that if LLAGNs have low Lorentz
factor (i.e. $\Gamma < 3$), the Doppler factor does not depend
strongly on the viewing angle and we should be able to see also
sources oriented at large $\theta$. The lack of raido cores in Type 2
LLAGN then may require alternative explanations such as free-free
absorption by a torus.

\subsection{History of emission}

From the result of the modelfit, small to moderate apparent velocities
($\la 0.2\,c$) are found for the four jet components. Under the
assumption of constant velocity, we derive that these components must
have been ejected from the core between 8.3 and 65.8 years before
epoch 2000.652 (see Column [6] in Table~\ref{tab4}).

However, these jet components are not to be confused with the hot
spots demarcating the end of the jet as found in more powerful CSOs
\citep{ows98,pec00,gir03}. Therefore, a kinematic estimate of the real
age of the source is not possible, and that of $S1$ can only be taken
as a lower limit. The low brightness and large size of $N2$, as well
as the non-detection of $N1$, suggest that components are continually
ejected from the core, but soon disrupt, without being able to travel
long distances and form kiloparsec-scale lobes.

We do not expect that this source will evolve into a kiloparsec-scale
radio galaxy, but rather that it will only periodically inflate
slowly, as visible in Fig.~\ref{fig5}. The relatively low velocity
jets discussed in \S~\ref{jets} cannot bore through the local ISM and
escape, as demonstrated by the lack of hot spots which are prevalent
in higher power CSOs. This behavior can be ascribed to a low power
central engine, which cannot create highly relativistic jets.

\begin{figure}
\plotone{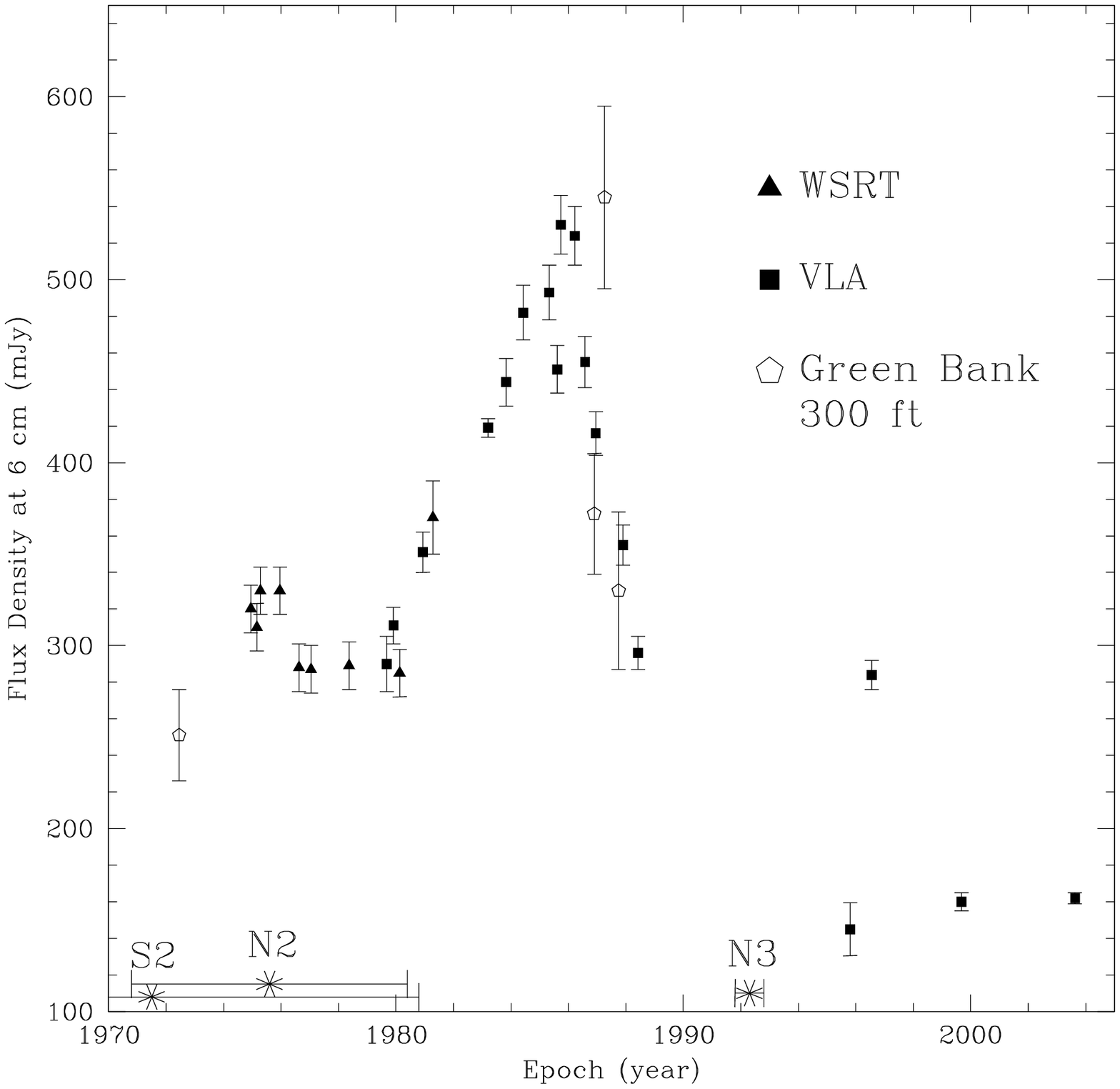}
\caption{The light curve for NGC~4278 at 6 cm. The asterisks mark the
  estimated epoch of emission of components $S2$, $N2$, and $N3$. The $S1$
  ejection is too far back in time to appear in this plot. Different
  symbols mark data taken with the WSRT \citep[filled triangles;
  see][]{eke83,sch83}, the VLA (filled squares; \citet{wro91} and
  references therein; \citet{nag01}; present work), and the Green Bank
  300~ft telescope \citep[empty
  pentagons;][]{sra75,lan90,bec91,gre91,gre96}.
  \label{lightcurve}}
\end{figure}

In Fig.~\ref{lightcurve} we plot the flux density history for NGC~4278
at 6 cm using data taken at the Westerbork Synthesis Radio Telescope,
the Green Bank 300~ft radio telescope and the VLA. A previous light
curve was published by \citet{wro91}, to which we add 12 points, from
observations obtained between 1972 and 2003. The light curve shows
that the radio source is variable, prone to both outbursts and low
states. A burst is certainly present around 1985, while in more recent
years the source has been showing less activity.

It is difficult to connect the burst with the ejection of new
components, both because of the uncertainties related to the age of a
single component, and the possible time lag between component ejection
and total flux density enhancement. It is clear however that the
source presents a high degree of variability, most likely related to
the presence of an active nucleus. This is an interesting result,
suggesting that repeated observations of other LLAGN are desirable,
given that few studies on this topic exist so far \citep{wro00,nag02}.

\subsection{The origin of energy in LLAGN}

Thanks to its proximity, NGC~4278 makes an excellent laboratory to
study the properties of LLAGNs and to explore the nature of the
physical processes taking place. In particular, there are three
important clues for solving the puzzle over the origin of the radio
emission: (1) the shape of the radio spectrum; (2) the parsec-scale
morphology, and (3) the brightness temperature of the VLBI core.

In 7 nearby galaxies studied by \citet{dim99,dim01}, ADAF models are
unsuccessful in fitting the high frequency radio spectra, and
synchrotron radiation from non-thermal particles in jets must
contribute significantly. Conversely, \citet{and04} have revealed in
six LLAGN VLBA spectral indexes that are remarkably flat or even
slowly rising up to 43 GHz. Although these cores are still too
luminous to be explained in terms of only radiation from an ADAF, the
jet properties remain to be determined and successfully imaged in
these sources.

In our case, the overall VLA spectrum can be fit by a power-law with
an intermediate index $\alpha = 0.54$ (see \S \ref{spix}). A possible
steepening occurs above 22 GHz, which argues against the presence of
an ADAF as significantly contributing to the total emissivity. A low
frequency flattening/turnover may be present around 408 MHz, with
$S_\mathrm{408\, MHz} =650$ mJy \citep{col72} and $S_\mathrm{365\, MHz} =
569$ mJy \citep{dou96}. Variability and resolution effects are also
important in the study of the spectrum of this source and other AGN --
for example, the flux density from the low brightness northwestern
region was missed by \citet{fal00}, resulting in a flatter
high-frequency spectral index. In any case, the radio spectrum between
151 MHz and 43 GHz is globally similar to that of more powerful radio
loud AGN, i.e. to that of synchrotron emission from relativistic
particles in a jet.

At higher frequencies, NGC~4278 possesses an optical central compact
core which lies nicely at the low power end of the correlation between
optical and radio core luminosity in FR~I radio galaxies
\citep{cap02}. This correlation can be interpreted in terms of a
common non-thermal synchrotron origin, with the radiation being
produced in relativistic jets \citep{cap99}.

The parsec-scale morphology revealed by our VLBA images is clearly in
agreement with this scenario, showing twin jets. This result and the
high brightness temperature $T_B = 1.5 \times 10^9$ K are in agreement
with similar findings from \citet{nag02}; therefore, we conclude that
the emission from NGC~4278 and from similar LLAGNs originates in radio
jets via the synchrotron process.

In this scenario, detection of polarized signals would be an
interesting signature of the synchrotron process; indeed, VLA
observations at 8.4 GHz have found weak linear polarization in
NGC~4278 at a level of 0.34\% and in other LLAGN, although the
detection rates and percentage of linear polarization in these sources
are lower than in more powerful AGNs \citep{bow02}. Our new
polarization data for NGC~4278 show that it is unpolarized on
milliarcsecond scales, as also found by \citet{bon04}.  This could be
due to limits in sensitivity. Since synchrotron radiation is expected
to have an intrinsically high percentage of linear polarization, one
needs to invoke a significant amount of Faraday depolarization, or
some other effect, to explain the low amount of linear polarization
observed. Note that low or absent linear polarization is a
characteristic of CSOs \citep{pec00,gir03}, although recent
observations have found exceptions to this rule \citep{gug04}.

Finally, it is interesting to discuss the position of the central
black hole responsible for the accretion and launch of the jets. Our
observations allow us to image the nuclear region of NGC~4278 on
scales of $\la 0.1$ pc (1 mas = 0.071 pc).  Based on the correlation
between the black hole mass and the central velocity dispersion, we
can estimate $M_\mathrm{BH}$ for NGC~4278. The relation takes the form
$\log (M_\mathrm{BH}/M_\sun) = 8.13 + 4.02 \log (\sigma/\sigma_0)$,
where $\sigma_0 = 200$~km~s$^{-1}$ \citep{tre02} and $\sigma = (258
\pm 15)$~km~s$^{-1}$ \citep{bar02}. It follows that $M_\mathrm{BH} =
3.7 \times 10^8 M_\sun$ and that the Schwarzschild radius $1
R_\mathrm{S} = 3.5 \times 10^{-5}$ pc; therefore, our images probe
scales of some $10^3 R_\mathrm{S}$.

\section{Conclusions}

We present new VLBA and VLA data for the nearby ($d=14.9$ Mpc) LLAGN
NGC~4278. Our new VLBA data show a two-sided emission on sub-parsec
scales in the form of twin jets emerging from a central compact
component ($T_B = 1.5 \times 10^9$ K), in a similar way to that seen
in more powerful radio loud AGNs. In agreement with this, the spectral
index distribution for the radio source reveals a flat spectrum core
region with steep spectrum jets on either side.

By comparison with previous observations, we discover proper motion
for components in both jets over a five years time baseline. We find
low apparent velocities ($\la 0.2\,c$) for the jet components and
estimate the epoch of their ejection as $10 - 100$ years prior to our
observations. Based on our analysis, we suggest that the north-west
side is the approaching side, and that the jets of NGC~4278 are mildly
relativistic with $\beta$ $\sim$ 0.75.

The central black hole in NGC~4278 is active and able to produce jets,
which are responsible for the bulk of the emission at radio to optical
frequencies in this LLAGN. However, the lifetime of the components of
$< 100$ years at the present epoch, combined with the lack of large
scale emission, suggests that the jets are disrupted before they reach
kiloparsec scales.

The study of the flux density history at 6 cm between 1972 and 2003
shows significant variability ($\ga 100\%$) on time scales of a few
years, which might be related to the ejection of new components. This
subject needs to be explored for other LLAGNs as well, as it can give
insights into the state of the central black hole in these sources.

\begin{acknowledgments}
We thank Joan Wrobel and an anonymous referee for helpful comments.
MG thanks the NRAO for hospitality during his visit to Socorro when
much of this work was accomplished. The National Radio Astronomy
Observatory is operated by Associated Universities, Inc., under
cooperative agreement with the National Science Foundation. This
research has made use of NASA's Astrophysics Data System Bibliographic
Services and of the NASA/IPAC Extragalactic Database (NED) which is
operated by the Jet Propulsion Laboratory, Caltech, under contract
with NASA.  This material is based in part upon work supported by the
Italian Ministry for University and Research (MIUR) under grant COFIN
2003-02-7534.

\end{acknowledgments}

\clearpage

\begin{deluxetable}{lccrrrrr}
\tabletypesize{\small}
\tablecaption{Image Parameters \label{tab1} }
\tablehead{
\colhead{} & \colhead{} & \colhead{beam} & \colhead{beam} & \colhead{}
& \colhead{peak} & \colhead{total} & \colhead{} \\
\colhead{Epoch} & \colhead{Freq.} & \colhead{size} &
\colhead{P.A.} & \colhead{noise} & \colhead{brightness} &
\colhead{flux density} & \colhead{$P$} \\
 & \colhead{(GHz)} & \colhead{(mas $\times$ mas)} &
\colhead{($^\circ$)} & \colhead{($\mu$Jy beam$^{-1}$)} &
\colhead{(mJy beam$^{-1}$)} & \colhead{(mJy)} & \colhead{(W Hz$^{-1}$)}
}
\startdata
1995.551 & 5.0 & 3.0 $\times$ 1.8 & $-25$ & 65 & 49.5 & $135 \pm 7$ & $3.57 \times 10^{21}$ \\
2000.652 & 5.0 & 3.0 $\times$ 1.7 & $-7$  & 45 & 40.5 & $120 \pm 6$ & $3.18 \times 10^{21}$ \\
2000.652 & 8.4 & 1.8 $\times$ 1.0 & $-4$  & 50 & 24.6 & $95 \pm 5$ & $2.52 \times 10^{21}$ \\
\enddata
\end{deluxetable}

\begin{deluxetable}{lrrrrrrrr}
\tablecaption{Results of the modelfit to epoch 2000.652
  \label{tab2} }
\tablehead{
\colhead{Component} & \colhead{$r$} & \colhead{$\theta$} &
\colhead{$a$} &\colhead{$b/a$}& \colhead{$\Phi$} &
\colhead{$S_\mathrm{5\, GHz}$} & \colhead{$S_\mathrm{8.4\, GHz}$} &
\colhead{$\alpha$} \\
       &   \colhead{(mas)}   &  \colhead{($^{\circ}$)}&     \colhead{(mas)}  &
& \colhead{($^{\circ}$)} &  \colhead{(mJy)} &  \colhead{(mJy)} &  }
\startdata
C  &  0.00 &     0.0 &  0.99 & 0.48 & $-$26.6 & 31.5 & 28.0 & 0.2 \\
S2 &  2.57 &   162.0 &  1.65 & 0.36 &    76.4 &  7.5 &  3.0 & 1.8 \\
S1 &  8.51 &   142.8 &  7.39 & 0.50 & $-$68.3 & 28.0 & 20.8 & 0.6 \\
N3 &  1.96 & $-$25.1 &  3.23 & 0.74 & $-$39.0 & 45.4 & 36.6 & 0.4 \\
N2 & 18.46 & $-$67.7 & 11.17 & 0.45 & $-$71.9 &  8.7 &  6.3 & 0.6 \\
\enddata
\tablecomments{Parameters of each Gaussian component of the model brightness
distribution: $r, \, \theta$, polar coordinates of
the center of the component relative to an arbitrary origin, with
polar angle measured from north through east; $a, \, b$, major and
minor axes of the FWHM contour; $\Phi$, position angle of the major
axis measured from north through east; $S_\mathrm{5\, GHz}, \,
S_\mathrm{8.4\, GHz}$, flux densities; $\alpha$, spectral index between 5
and 8.4 GHz.}
\end{deluxetable}

\begin{deluxetable}{lrrrrrr}
\tablecaption{Results of the modelfit to epoch 1995.551
  \label{tab3} }
\tablehead{
\colhead{Component} & \colhead{$r$} & \colhead{$\theta$} &\colhead{$a$} &\colhead{$b/a$}& \colhead{$\Phi$} & \colhead{$S_\mathrm{5\, GHz}$} \\
       &    \colhead{(mas)}   &  \colhead{($^{\circ}$)}&     \colhead{(mas)}  &
& \colhead{($^{\circ}$)} &  \colhead{(mJy)}  }
\startdata
C  &  0.00 &     0.0 &   0.99 & 0.48 & $-$26.6 & 26.7   \\
S2 &  2.13 &   164.8 &   1.65 & 0.36 &    76.4 & 10.8   \\
S1 &  7.85 &   142.6 &   6.00 & 0.58 & $-$71.1 & 32.8   \\
N3 &  1.32 &    12.1 &   2.86 & 0.46 & $-$23.4 & 52.5   \\
N2 & 14.80 & $-$64.7 &  12.51 & 0.32 & $-$67.5 & 13.5   \\
\enddata
\tablecomments{Parameters of each Gaussian component of the model brightness
distribution: $r, \, \theta$, polar coordinates of
the center of the component relative to an arbitrary origin, with
polar angle measured from north through east; $a, \, b$, major and
minor axes of the FWHM contour; $\Phi$, position angle of the major
axis measured from north through east; $S_\mathrm{5\, GHz}$, flux
density at 5 GHz.}
\end{deluxetable}

\begin{deluxetable}{lrrrrrrr}
\tablecaption{Component motion at 5 GHz
  \label{tab4} }
\tablehead{
\colhead{Component} & \colhead{$\Delta r$} & \colhead{$\rho$} &\colhead{$\mu$}
&\colhead{$\beta_\mathrm{app}$} & \colhead{age} \\
       &    \colhead{(mas)}   &  \colhead{($^{\circ}$)}&
\colhead{(mas~yr$^{-1}$)}  & \colhead{($v_\mathrm{app}/c$)} & \colhead{(yrs)} }
\startdata
C  & \multicolumn{5}{l}{reference} \\
S2 & $0.45 \pm 0.14$ & 148.7   & $0.088 \pm 0.027$ & $0.020 \pm 0.006$ & $29.1 
\pm 9.3$ \\			    
S1 & $0.66 \pm 0.12$ & 145.1   & $0.129 \pm 0.024$ & $0.030 \pm 0.006$ & $65.8
\pm 12.4$ \\
N3 & $1.21 \pm 0.09$ & $-$66.3 & $0.237 \pm 0.018$ & $0.055 \pm 0.004$ & $8.3
\pm 0.5$ \\
N2 & $3.76 \pm 0.65$ & $-$79.5 & $0.737 \pm 0.128$ & $0.171 \pm 0.030$ & $25.0 
\pm 4.8$ \\
\enddata
\end{deluxetable}

\begin{deluxetable}{cr}
\tablecaption{VLA Flux Densities \label{tabvla} }
\tablehead{
\colhead{Frequency} & \colhead{Flux Density} \\
\colhead{(GHz)} & \colhead{(mJy)}
}
\startdata
1.4 & $331 \pm 7$ \\
5.0 & $162 \pm 3$ \\
8.4 & $114 \pm 2$ \\
15  & $99 \pm 5$ \\
22  & $73 \pm 7$ \\
43  & $<50$ \\
\enddata
\end{deluxetable}

\clearpage

\end{document}